\documentclass[journal]{IEEEtran}

\IEEEoverridecommandlockouts
\usepackage{cite}
\usepackage{amsmath,amssymb,amsfonts}
\usepackage{graphicx}
\usepackage{textcomp}
\usepackage{xcolor}
\def\BibTeX{{\rm B\kern-.05em{\sc i\kern-.025em b}\kern-.08em
    T\kern-.1667em\lower.7ex\hbox{E}\kern-.125emX}}
\usepackage{mathtools}
\usepackage{amsthm}
\usepackage{stfloats}
\usepackage{epstopdf}
\usepackage[linesnumbered,ruled,vlined]{algorithm2e}

\usepackage{subcaption}
\usepackage{algcompatible}
\usepackage{ragged2e}
\usepackage{float}
\usepackage[normalem]{ulem}	
\usepackage{soul}

\begin{document}

\title{Integrated Sensing and Communications with Affine Frequency Division Multiplexing}

\author{
    \IEEEauthorblockN{Ali Bemani\IEEEauthorrefmark{1}, Nassar Ksairi\IEEEauthorrefmark{1}, and Marios Kountouris\IEEEauthorrefmark{2}}
    
    \IEEEauthorblockA{\IEEEauthorrefmark{1}Mathematical and Algorithmic Sciences Lab, Huawei France R\&D, Paris, France,
    \\\{ali.bemani,nassar.ksairi@huawei.com}\}

    \IEEEauthorblockA{\IEEEauthorrefmark{2}Communication Systems Department, EURECOM, Sophia Antipolis, France
    \\ marios.kountouris@eurecom.fr}
\vspace{-0mm}
}

\maketitle
\begin{abstract}
Integrated sensing and communications (ISAC) is regarded as a key technology in next-generation (6G) mobile communication systems. Affine frequency division multiplexing (AFDM) is a recently proposed waveform that achieves optimal diversity gain in high mobility scenarios and has appealing properties in high-frequency communication. In this letter, we present an AFDM-based ISAC system. We first show that in order to identify all delay and Doppler components associated with the propagation medium, either the full AFDM signal or only its pilot part consisting of one discrete affine Fourier transform (DAFT) domain symbol and its guard interval can be used. Our results show that using one pilot symbol achieves almost the same sensing performance as using the entire AFDM frame. Furthermore, due to the chirp nature of AFDM, sensing with one pilot provides a unique feature allowing for simple self-interference cancellation, thus avoiding the need for expensive full duplex methods.
\end{abstract}
\begin{IEEEkeywords}
AFDM, ISAC, affine Fourier transform, chirps, doubly dispersive channels.
\end{IEEEkeywords}
\vspace{-0mm}
\section{Introduction}
Wireless communications and radar sensing have historically been developed as separate fields, each with its own specific requirements and operating frequency bands. In recent years, the consolidation of these two domains into an integrated sensing and communication (ISAC) paradigm has garnered significant attention, largely owing to its promise of lower power usage, enhanced spectral efficiency, and decreased hardware costs \cite{liu2020joint}.

The design of a dual-functional waveform, which achieves integration gain by sharing signaling resources for sensing and communication, is crucial in ISAC systems. Orthogonal frequency division multiplexing (OFDM) has been extensively studied in ISAC waveform design and signal processing methods 
\cite{geng2023novel} however, it suffers from inter-carrier interference and consequent communication performance deterioration in high mobility scenarios. Orthogonal chirp division multiplexing (OCDM) \cite{ouyang2016orthogonal} is an alternative scheme based on the discrete Fresnel transform, which has a lower bit error rate (BER) than OFDM \cite{Oliveira2020}, at the cost of higher computational complexity. Nevertheless, OCDM cannot achieve the optimal diversity order of linear time-varying (LTV) channels \cite{bemani2023affine_TWC}, resulting as well in higher hardware complexity. Another modulation technique, orthogonal time frequency space (OTFS), has been investigated for ISAC, e.g., in \cite{gaudio2020effectiveness}, due to its inherent link to the delay-Doppler domain. Nonetheless, all aforementioned methods rely on the strong assumption of ideal self-interference cancellation (SIC), which in turn necessitates costly full duplex SIC methods.

A recently proposed technique that can achieve robust communication performance in high mobility scenarios is Affine Frequency Division Multiplexing (AFDM) \cite{bemani2021afdm,bemani2023affine_TWC}. AFDM employs multiple orthogonal chirps generated using the discrete affine Fourier transform (DAFT). With chirp parameters adapted to the channel characteristics, AFDM can reconstruct a delay-Doppler representation of the channel achieving full diversity on doubly dispersive channels. In comparison with OTFS, AFDM has comparable BER performance but with the advantage of requiring less channel estimation overhead \cite{bemani2023affine_TWC}. This makes AFDM a potential candidate waveform for ISAC \cite{wang2022towards}. A first AFDM-based ISAC scheme with a sensing approach that uses the full AFDM signal (both its data and pilot parts) is proposed in \cite{Ni2022}, showing very good sensing performance even in large Doppler scenarios, however only under the restrictive assumption of target ranges corresponding to integer-valued delays.

In this letter, we propose a novel ISAC scheme leveraging AFDM and we analyze its performance under both fractional delay and Doppler shifts, i.e., without requiring the restrictive integer-delays assumption of \cite{Ni2022}. We show that sensing in AFDM-based ISAC can be done using either the whole frame as in \cite{Ni2022}, or one pilot symbol, motivated from \cite{bemani2023affine_TWC} in which as few as one DAFT domain symbol used as a pilot can yield - when appended with a sufficient number of zero guard samples - the possibility to identify all the delay and Doppler components associated with the propagation medium. Evidently, this is relevant for sensing and radar applications since the delay-Doppler representation of the wireless channel associated with the round-trip propagation from the wireless transmitter to the targets in its vicinity and back to the transmitter translates into range-velocity information about those targets. Moreover, we show that by sensing with one DAFT domain pilot rather than the whole signal, a low-complexity SIC can be implemented even when this pilot is multiplexed with data and possibly other pilots.

\section{System Model}
\begin{figure}
    \centering
    \includegraphics[scale=.8]{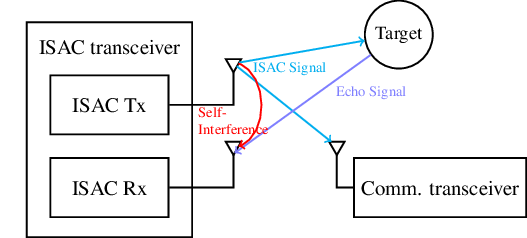}
    \caption{Monostatic ISAC system model}
    \label{fig:ISAC_diagram}
\end{figure}

The considered ISAC system is shown in Fig. \ref{fig:ISAC_diagram} and consists of an AFDM-based ISAC transceiver and an AFDM receiver. The ISAC transceiver is equipped with a single antenna monostatic radar and uses the same AFDM waveform for both communication and sensing. The transceiver conveys messages to the receiver while estimating parameters related to the targets using the reflected signal from the targets.
\subsection{AFDM-based ISAC signal model}
\begin{figure}
    \centering
    \includegraphics[scale=.57]{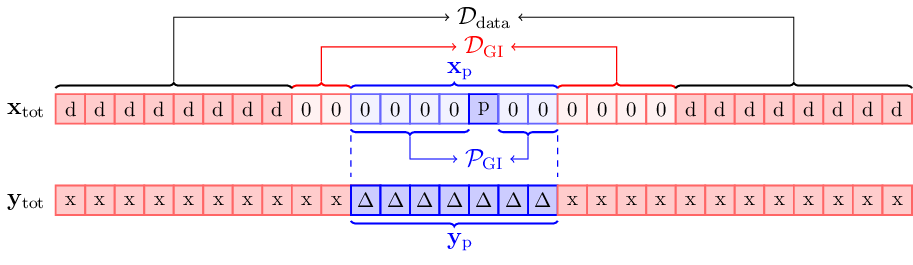}
    \caption{Transmitted and received AFDM frame}
    \label{fig:Frame}
\end{figure}

\begin{figure}
    \centering
    \includegraphics[scale=.50]{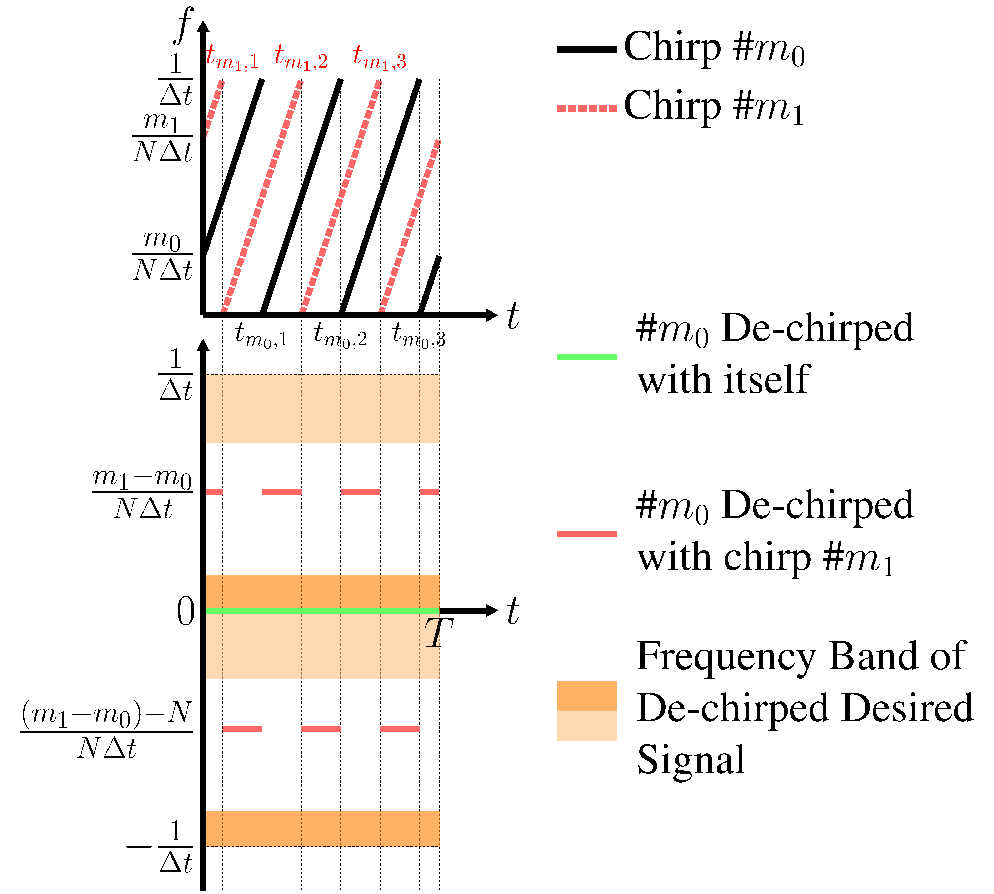}
    \caption{Time-frequency representation of two AFDM chirps (the pilot, $m_0$, and one chirp, $m_1$, from outside the pilot guard interval) at the transmitter (on top) and after analog dechirping at the radar receiver (on bottom)}
    \label{fig:Chirp_corr}
\end{figure}

The embedded-pilot scheme proposed in \cite{bemani2023affine_TWC} is next considered for AFDM-based ISAC. An $N$-long DAFT domain frame from this scheme $x[m],m=0\cdots N-1$ is illustrated in Fig. \ref{fig:Frame}. In this frame, the embedded pilot symbol $x_{\rm{pilot}}$ is transmitted at index $m_0$ while $\mathcal{P}_{\rm{GI}}$ is the set of null-guard indices useful for pilot based channel estimation and sensing while $\mathcal{D}_{\rm{GI}}$ is the set of null-guard indices that guarantee pilot-data orthogonality. The set of indices $\mathcal{D}_{\mathrm{data}} \triangleq \{0, ..., N-1\} \backslash (\mathcal{P}_{\rm{GI}} \cup \mathcal{D}_{\rm{GI}} \cup \{m_0\})$ is used for transmitting data symbols $\left\{x_m^{\rm{data}}\right\}_{m\in\mathcal{D}_{\mathrm{data}}}$.
The way the guard intervals are chosen to guarantee pilot-data orthogonality is detailed after \eqref{eq:y_received}.
Finally, we set $\mathbb{E}[|x_{\rm{pilot}}|^2]=|\mathcal{P}_{\rm{GI}} \cup \mathcal{D}_{\rm{GI}}|\mathbb{E}[|x_m^{\rm{data}}|^2]$ because the DAFT domain null interval allows for boosting the pilot power without resulting in an increase in the average transmit power. This hints at a trade-off between sensing signal power (proportional to pilot overhead  $\triangleq\frac{|\mathcal{P}_{\rm{GI}} \cup \mathcal{D}_{\rm{GI}}|}{N}$) and communication spectral efficiency (proportional to $\left|\mathcal{D}_{\rm data}\right|$).
AFDM transmitter maps $\{x[m]\}_{m=0\cdots N-1}$ to $\{s[n]\}_{m=0\cdots N-1}$ using the inverse DAFT (IDAFT) with parameters $(c_1,c_2)$ \cite{bemani2023affine_TWC} as follows:
{\small{\begin{equation}
\label{eq:mod}
    s[n] = \frac{1}{\sqrt{N}}\sum_{m = 0}^{N-1}x[m]e^{\imath2\pi (c_2m^2+{\frac{1}{N}}mn+c_1n^2)}, n= 0 \cdots N-1
\end{equation}}}Next, the sequence $\{s[n]\}_{n=0\cdots N-1}$ is appended with a chirp-periodic prefix (CPP) \cite{bemani2023affine_TWC} by transmitting samples $s[N+n]e^{-\imath2\pi c_1(N^2+2Nn)}$ at $n = -M, -M+1, ..., -1$. Here, $M$ denotes the CPP duration that is supposed to be larger than both the maximum delay of the communication channel and the maximum round trip delay in samples of the radar targets. The CPP simplifies to a cyclic prefix (CP) when $2c_1N\triangleq C$ is an integer and $N$ is even \cite{bemani2023affine_TWC}. We define $s(t)$ as the continuous-time version of $s[n]$ given by
\begin{equation}
\label{eq:mod_con}
    s(t) = \frac{1}{\sqrt{T}}\sum_{m = 0}^{N-1}x[m]e^{\imath2\pi (c_2m^2+\Phi_m(t))}, \quad 0 \leq t \leq T,
\end{equation}
where $c'_1 = \frac{c_1}{\Delta t^2}$, $\frac{1}{\Delta t}$ is the sampling rate, $T=N\Delta t$ and $2\pi\Phi_m(t)$ is the $m$-th chirp instantaneous phase. Due to the frequency wrapping property of AFDM chirp carriers shown in Fig. \ref{fig:Chirp_corr}, $\Phi_m(t)$ should be defined piece-wise on intervals corresponding to the partition $\{t_{m,q}\}_{q = 0, \cdots, C}$ of $[0, T)$ where $t_{m, 0} = 0$ and $t_{m, q} = \frac{(N-m)}{2Nc_1}\Delta t + \frac{q-1}{2c_1}\Delta t$. On each interval $[t_{m, q}, t_{m, q+1})$, $\Phi_m(t)$ is defined such that 1) its derivative coincides with the instantaneous frequency given in Fig. \ref{fig:Chirp_corr} and 2) $\sqrt{\Delta t}s(n\Delta t) = s[n]$ for all $n\in\{0,\ldots N-1\}$. Using these two conditions we get that $\Phi_m(t) = c_1't^2 + \frac{m}{T}t + \alpha_m(t)$  where $\alpha_m(t) = - \frac{q}{\Delta t}t$ for $t\in[t_{m, q}, t_{m, q+1})$. Finally, $s_{\rm{CPP}}(t)$ is defined as the chirp-periodic continuous-time signal whose restriction on $[0,T]$ is $s(t)$.
In the case of $P$ point targets, the wireless link between the ISAC transmitter and receiver is represented by a $(P+1)$-tap time-frequency (doubly) selective channel 
\begin{equation}
    {h}(t, \tau) = \sum _{i=0}^{P} h_{i}e^{-\imath2\pi f_it}\delta(\tau - \tau_i)\label{eq:ch_model}
\end{equation}
where $i = 0$ stands for the direct path $h(t, 0) = h_0$ between the ISAC transmitter and receiver (the direct path has zero delay and Doppler shift). From now on we assume $h_0 = 0$, a justified assumption if some effective SIC (such as the scheme of Section \ref{sec:SIC}) is applied. $h_i, f_i$ and $\tau_i$ are the complex gain, Doppler shift, and the delay associated with the $i$-th target, respectively. In this paper, we assume that the number of targets is known. Moreover, we assume that the range of the delay and Doppler shift is $\left[0,\tau_{\max}\right]$ and $\left[-f_{\max},f_{\max}\right]$, respectively, where $\tau_{\max}$ represents the maximum delay in seconds and $f_{\max}$ is the maximum Doppler shift in Hz. Let $c$, $f_c$, $v_i$, $r_i$ denote the speed of light (m/s), carrier frequency (Hz), corresponding velocity (m/s), and range (m) associated with the $i$-th target, respectively. Thus, the relative range and velocity of the $i$-th target are respectively $r_i = c\cdot\frac{\tau_i}{2}$ and $v_i = c\cdot\frac{f_i}{2f_c}$. The delay and Doppler in samples associated with the $i$-th path are denoted $l_i$ and $k_i$, respectively, so that
\begin{equation}
l_i + \iota_i = \tau_i/\Delta t, \quad k_i +\kappa_i = Tf_i,
\end{equation}
where $k_i$ denotes the integer part of the normalized Doppler shift for the $i$-th path, and $\frac{-1}{2} < \kappa_i \leq \frac{1}{2}$ represents the corresponding fractional Doppler shift. Notations $l_i$ and $\frac{-1}{2} < \iota_i \leq \frac{1}{2}$ are similarly defined for delay shifts. Normalized Doppler and delay shifts satisfy $k_i + \kappa_i \in \left[-k_{\max}, k_{\max}\right]$ where $k_{\max}\triangleq\lceil Tf_{\rm{max}}\rceil$ is the maximum Doppler shift in samples and $l_i + \iota_i \in \left[0,l_{\max}\right]$ where $l_{\max}\triangleq\lceil \tau_{\rm{max}}/\Delta t\rceil$ is the maximum delay in samples. After transmission over the channel, the received signal at the ISAC receiver is written as 
\begin{equation}
\label{eq:rt}
      r(t) =\sum_{i = 1}^{P}s_{\rm{CPP}}(t-\tau_i)h(t, \tau_i) + w(t),
\end{equation}
where $w(t)$ is the additive white noise.  By sampling every $\Delta t$ seconds, removing the CPP and applying DAFT, output symbols are obtained after some simplification as
{\small
\begin{equation}
\begin{aligned}
    &y[m] = \frac{1}{\sqrt{N}}\sum_{n = 0}^{N-1}r[n]e^{-\imath2\pi (c_2m^2+{1\over N}mn+c_1n^2)}\\
    &=\frac{1}{N}\sum_{i = 1}^{P}\sum_{m' = 0}^{N-1}h_ix[m']e^{\imath2\pi ( c_1(l_i + \iota_i)^2 - c_2(m^2 - m'^2)-\frac{(l_i+\iota_i)m'}{N})}\\
    & \times \underbrace{\sum_{n = 0}^{N-1}e^{\imath\frac{2\pi}{N}(m' - (m + l^{\rm{eq}}_i))n}e^{\imath2\pi\iota_i (\sum_{q = 0}^{C} qI_{\boldsymbol{\mathcal{L}}_{m,q}}((n - (l_i + \iota_i))_N)) }}_{\mathcal{F}_{i}(m, m')} + {w}[m]
    \label{eq:y_received}
\end{aligned}
\end{equation}
}where $r[n] = \sqrt{\Delta t}r(t)|_{t = n\Delta t}$, $l^{\rm{eq}}_i \triangleq (k_i +\kappa_i)+2Nc_1(l_i+\iota_i)$ (`$\mathrm{eq}$' stands for ``equivalent delay''), $I_{\boldsymbol{\mathcal{L}}_{m,q}}$ is the indicator function of the set $\boldsymbol{\mathcal{L}}_{m,q} \triangleq \{n_{m, q} + 1,\cdots n_{m, q+1}\}$ with $n_{m, q} \triangleq \lfloor \frac{t_{m,q}}{\Delta t}\rfloor$ and  ${{w}}[m]$ is an i.i.d. noise  with $\sim\mathcal{CN}\left(0,N_0\right)$. It is pertinent to highlight that ${\mathcal{F}_{i}(m, m')}$ in \eqref{eq:y_received} simplifies to $\frac {e^{\imath {2\pi } (m'-m-l^{\rm{eq}}_p) }\!-\!1}{e^{\imath \frac {2\pi }{N} (m'-m-l^{\rm{eq}}_p)}\!-\!1}$ for zero fractional delay, and to $N\delta(m' - (m + l^{\rm{eq}}_p)_N)$ when both the normalized delay and Doppler shifts are integers. With a pilot symbol at DAFT index $m_0$ as in Fig. \ref{fig:Frame}, we showed in \cite{bemani2023affine_TWC} that an input-output relation as the one in \eqref{eq:y_received} leads to the received samples related to the above pilot symbol being significant only in the following interval
\begin{equation}
   m_0 - (k_f + k_{\max}) - 2Nc_1l_{\max} \leq m \leq m_0 + (k_f + k_{\max}),\label{eq:guard_intveral}
\end{equation}
provided $k_f$ is chosen large enough (for $\left|\mathcal{F}_{i}(m, m_0)\right|$ to be small enough outside the above interval) and where in the case of negative indices modulo $N$ operation applies. We thus set the guard index set $\mathcal{P}_{\rm{GI}}$ to be equal to the interval in \eqref{eq:guard_intveral}. The set $\mathcal{D}_{\rm{GI}}$ is chosen to be of the same length. 

\textit{Remark on AFDM Parameter Settings:} The performance of AFDM-based sensing significantly depends on the value of parameters \( c_1 \). In our system, setting \( c_1 \) to \( \frac{2(k_f + k_{\max})}{2N} \) as explained in \cite{bemani2023affine_TWC} is key, ensuring a full delay-Doppler channel representation and facilitating efficient target separation within the DAFT domain with as few as one DAFT domain pilot symbol. Moreover, this setting is essential for enabling the low-complexity SIC scheme presented in Section \ref{sec:SIC}.

\section{AFDM-based Sensing}\label{sec:estimation}
The radar input-output relation \eqref{eq:y_received} writes in matrix form as
\begin{equation}
    {\mathbf y} = \sum_{i = 1}^{P}h_i\mathbf{H}_i(\tau_i,f_i)\mathbf{x} + {\mathbf w}
    \:.\label{eq_rec}
\end{equation}
Here, $\mathbf{y}$ is a general notation we use to designate either 1) the $N$-long vector composed of all the received DAFT domain samples corresponding to the whole vector of transmitted symbols, in which case $\mathbf{y}=\mathbf{y}_{\rm tot}$ and $\mathbf{x}=\mathbf{x}_{\rm tot}$, or 2) the $\left(\left|\mathcal{P}_{\rm GI}\right|+1\right)$-long vector composed of the received samples corresponding to the single pilot symbol, $x_{\rm{pilot}}$, and its guard samples as depicted in blue in Fig. \ref{fig:Frame}, in which case $\mathbf{y}=\mathbf{y}_{\rm p}$ and $\mathbf{x}=\mathbf{x}_{\rm p}$. Matrices $\mathbf{H}_i$ are defined similarly and can be deduced by comparing \eqref{eq_rec} and \eqref{eq:y_received}. Target range and velocity estimation can certainly be done utilizing all received DAFT domain samples $\mathbf{y}_{\rm tot}$ because in our monostatic radar setting the entire frame $\mathbf{x}_{\rm tot}$ is available for sensing. An alternative approach involves doing the estimation using only the segment $\mathbf{y}_{\rm p}$ of the received signal. In both cases, an approximate maximum likelihood (ML) estimate of $\boldsymbol{\tau} = [\tau_1, ..., \tau_{P}]$ and  $\boldsymbol{f} = [f_1, ..., f_{P}]$ can be obtained \cite{bemani2023affine_TWC} by solving    
\begin{equation} 
[\hat{\boldsymbol{\tau}}, \hat{\boldsymbol{f}}] = {\rm{arg}} \max _{[\boldsymbol{\tau}, \boldsymbol{f}] \in \mathbb{R}^P\times \mathbb{R}^P}\sum_{i=1}^{P}{|\mathbf{x}^H{\mathbf{H}}_i^H(\tau_i,f_i){{\mathbf y}}|^2}.
\label{eq:max_ML_approx}
\end{equation}
Next, the range and velocity estimates can be obtained as $\hat{r}_i = c\cdot\frac{\hat{\tau}_i}{2}$ and $\hat{v}_i = c\cdot\frac{\hat{f}_i}{2f_c}$. A practical approach to solve \eqref{eq:max_ML_approx} is a refined-grid search with steps $\Delta_\tau, \Delta_f \in (0,1]$ for the delay and Doppler shifts, respectively, having a complexity of $O(P(l_{\rm max}+1)(2k_{\max} + 1)\frac{Q^2}{\Delta_f\Delta_\tau}))$ where $Q = N$ when sensing is done using the whole frame $\mathbf{y}=\mathbf{y}_{\rm tot}$ and $Q = \left(\left|\mathcal{P}_{\rm GI}\right|+1\right)$ when sensing is done using only its pilot part $\mathbf{y}=\mathbf{y}_{\rm p}$. Using only a subset of the time-frequency resources of ISAC signals for sensing has already been proposed, e.g., \cite{geng2023novel}, to reduce computational complexity ($\left(\left|\mathcal{P}_{\rm GI}\right|+1\right) < N$) or to decouple the problems of beamforming design for sensing and for communications. In the subsequent subsection, we show that AFDM sensing using only $\mathbf{y}=\mathbf{y}_{\rm p}$ has an additional benefit in terms of SIC simplicity. While a trade-off naturally arises in pilot-based sensing between pilot overhead and the mean-squared error (MSE) performance associated with $\hat{r}_i$ and $\hat{v}_i$, results of Section \ref{sec:sim} demonstrate that the effective resolution that can be achieved when solving \eqref{eq:max_ML_approx} is only marginally improved when using the whole frame.
\subsection{SIC for AFDM-based sensing}\label{sec:SIC}
A significant challenge in the monostatic radar configuration assumed in this work arises from the direct path $h(t,0)=h_0$ between the co-located ISAC transmitter and receiver, as depicted in Fig. \ref{fig:ISAC_diagram}. This path introduces substantial self-interference (SI), which not only interferes with the desired signal reflected by the targets but is also orders of magnitude stronger. This SI can severely impact the dynamic range requirements for the analog-to-digital converter (ADC) in the ISAC receiver, leading to a degradation in radar sensing performance, especially with practical ADCs. Therefore, implementing an effective SIC solution prior to the ADC stage becomes a crucial aspect of our system design to mitigate these challenges.

We now show that the assumption $h_0=0$ we made following \eqref{eq:ch_model} is justifiable when sensing is performed using only the AFDM pilot signal i.e., $\mathbf{y}_{\rm p}$, because in that case SIC can be achieved with simple analog dechirping and filtering. This is a major advantage over the AFDM-based ISAC in \cite{Ni2022} and OFDM- and OTFS-based ISAC which all need costly full-duplex SIC solutions. It also has a significant advantage over OCDM-based ISAC, as we show below. Dechirping $r(t)$ as given by \eqref{eq:rt} using the $m_0$-th chirp $R(t) \triangleq \frac{1}{\sqrt{T}}e^{\imath2\pi \Phi_{m_0}(t)}$ as reference yields
{\small
\begin{equation}
\begin{aligned}
    &r(t)R^*(t) = w(t)R^*(t)\\[-2.3em]
          & +\underbrace{s_{\rm{CPP}}(t)h(t, 0)R^*(t)}_{I(t):\text{dechirped SI signal}} + \overbrace{\sum_{i = 1}^{P}s_{\rm{CPP}}(t-\tau_i)h(t, \tau_i)R^*(t).}^{E(t) : \text{dechirped echo signal}}
\end{aligned}
\end{equation}}
Defining $\xi_{m,m_0}(t) \triangleq{e^{\imath2\pi ({\frac{(m - m_0)}{T}})t}e^{\imath2\pi(\alpha_m(t) - \alpha_{m_0}(t) )}}$ and ${\zeta_{i,m,m_0}(t)} \triangleq{e^{\imath2\pi({-2c_1\tau_i - f_i + \frac{(m - m_0)}{T}})t}e^{\imath2\pi(\alpha_{m}((t-\tau_i)_T) - \alpha_{m_0}(t))}}$, the dechirped SI and echo signal on $[0,T]$ are given as
{\small
\begin{equation}
    I(t) = \frac{1}{T}h_0x'_{\rm{pilot}}
    + \frac{1}{T}h_0\sum_{m\in \mathcal{D}_{\mathrm{data}}}x'[m]{\xi_{m,m_0}(t)}\left(\text{SI}\right),
    \label{eq:SI_corr}
\end{equation}
\begin{equation}
    \begin{aligned}
         &E(t) = \sum_{i = 1}^{P} h'_ix'_{\rm{pilot}}e^{-\imath2\pi \frac{\tau_im_0}{T}}{\zeta_{i,m_0,m_0}(t)}\left(\text{desired}\right)  \\[-1mm]
       &+ \sum_{i = 1}^{P}h'_i\sum_{m\in \mathcal{D}_{\mathrm{data}}}x'[m]e^{-\imath2\pi{\frac{\tau_im}{T}}}{\zeta_{i,m,m_0}(t)} , \left(\text{data echos}\right)
       \end{aligned}
    \label{eq:DS_corr}
\end{equation}}
where $x'[m] \triangleq x[m]e^{\imath2\pi c_2m^2}$, $x'_{\rm{pilot}} \triangleq e^{\imath2\pi c_2m_0^2}x_{\rm{pilot}}$, $h'_i \triangleq \frac{1}{T}h_ie^{\imath2\pi c'_1\tau_i^2}$.
To gain insight into \eqref{eq:SI_corr} and \eqref{eq:DS_corr}, we refer to Fig. \ref{fig:Chirp_corr}, which shows the time-frequency representation of an AFDM signal with two active chirp carriers, namely $m_0$ and $m_1$, where $m_0$ is the sensing pilot while $m_1$ carries data and of the output of analog dechirping done using $m_0$ as reference chirp. In \eqref{eq:SI_corr}, the first term, which corresponds to the pilot symbol, is a direct current (DC) component (shown in green in Fig. \ref{fig:Chirp_corr}) and can be eliminated with a DC blocking module. The second term, which is the part of the dechirped SI signal that is related to the data symbols, is a weighted sum of complex exponentials $\left\{\xi_{m,m_0}(t)\right\}_{m\in\mathcal{D}_{\rm data}}$. The second term of \eqref{eq:DS_corr}, i.e., the data part of the dechirped echo signal, is a weighted sum of functions $\left\{\zeta_{i,m,m_0}(t)\right\}_{m\in\mathcal{D}_{\rm data}}$ and can be written thanks to the approximation in \eqref{eq:guard_intveral} as a sum of complex exponentials $\left\{\xi_{m,m_0}(t)\right\}_{m\in\mathcal{D}_{\rm data}\cup\mathcal{D}_{\rm GI}}$. Consequently, the part of the dechirped signal related to data (whether from SI or echos) occupies in frequency a band defined by $\mathcal{D}_{\rm data}\cup\mathcal{D}_{\rm GI}$ and shown in white in Fig. \ref{fig:Chirp_corr}.
Note that this band is composite (made up of two separate parts) because the instantaneous frequency of $\xi_{m,m_0}(t)$ has jumps, as illustrated with solid red lines in the figure, because $\alpha_{m}(t) - \alpha_{m_0}(t)$ can take different values $0,-\frac{1}{\Delta t}t$ or $\frac{1}{\Delta t}t$ depending on $m$ and $ m_0$.

\begin{figure}
    \centering
    \includegraphics[width = .48\textwidth]{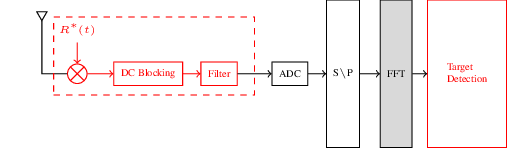}
    \caption{AFDM sensing receiver}
    \label{fig:ISAC_Rx}
\end{figure}

The first term of \eqref{eq:DS_corr} is the dechirped desired signal containing the useful delay-Doppler information of the targets and can be utilized for range-velocity estimation. Using the approximation in \eqref{eq:guard_intveral}, each function $\zeta_{i,m_0,m_0}(t)$ in \eqref{eq:DS_corr} can be written as a sum of complex exponentials $\left\{\xi_{m,m_0}(t)\right\}_{m\in\mathcal{P}_{\rm GI}}$. Consequently, the useful part of the dechirped signal occupies in frequency a band defined by $\mathcal{P}_{\rm GI}$ and shown with two shades of orange in Fig. \ref{fig:Chirp_corr}. As was the case for data, this band is composite (made up of three separate parts) because of the jumps in the instantaneous frequency of $\xi_{m,m_0}(t)$. More importantly, it does not overlap with the spectral components of any data chirp $m\in\mathcal{D}_{\rm data}\cup\mathcal{D}_{\rm GI}$ (the white region). SIC for AFDM-based sensing thus boils down to applying an analog filter with three pass bands: one central band around the zero frequency of a width equal to $\frac{\left|\mathcal{P}_{\rm GI}\right|+1}{T}$ Hz (illustrated with two shades of orange in Fig. \ref{fig:Chirp_corr}), the other two, adjacent to the $\frac{1}{\Delta t}$, and $-\frac{1}{\Delta t}$ frequencies mirroring the positive and the negative valued portions of the central band. Note that after analog-to-digital conversion at a sampling rate $\frac{1}{\Delta t}$ (or any rational fraction of $\frac{1}{\Delta t}$ larger than $\frac{\left|\mathcal{P}_{\rm GI}\right|+1}{T}$), the dechirped desired signal occupies the equivalent (in digital frequencies) of only the central part of the orange region due to the disappearance of the frequency jumps thanks to spectrum folding. The resulting SI-free and frequency-jumps-free discrete-time signal can be used to get the vector $\mathbf{y}_{\rm p}$ needed for sensing based on \eqref{eq:max_ML_approx} by passing the signal to an $N$-point FFT (see Fig. \ref{fig:ISAC_Rx}) and keeping $\left|\mathcal{P}_{\rm GI}\right|+1$ of its output samples.

\textit{Remark on the complexity of SIC for AFDM-based ISAC:} To summarize, SIC in AFDM-based monostatic sensing requires only analog dechirping followed by analog filtering. This is to be contrasted with OFDM and OTFS for which only costly full-duplex SIC methods are available. Also, the fact that the IS-free signal following analog dechirping and filtering in AFDM-based sensing only occupies a small portion equal to $\frac{\left|\mathcal{P}_{\rm GI}\right|+1}{T}$ of the system bandwidth $\frac{1}{\Delta t}$ means that there is no need for high-rate ADC, as would be needed for OFDM and OTFS radar receivers. The low-complexity/cost SIC of AFDM is also to be contrasted with the higher complexity of its OCDM counterpart. Indeed, sensing based on an OCDM frame of the same size as an AFDM frame from our setting would require approximately $\frac{l_{\max}(2k_{\max}+1)}{l_{\max}+2k_{\max}}$ chirp pilots \cite{benzine2023affine}, each with its own guard interval, because OCDM does not achieve the full diversity of LTV channels as opposed to AFDM. Therefore, SIC would require multiple RF chains, with a number equal to that of OCDM chirp pilots, resulting thus in a corresponding increase in hardware complexity.

\section{Simulation results}
\label{sec:sim}
We use the simulation setup parameters of Table \ref{tab:setup} to evaluate the performance of  the two variants of the proposed ISAC scheme in terms of range and velocity estimation and we compare it with OTFS- and OCDM-based ISAC.
\begin{table}
\centering

\caption{Simulations setting}
\begin{tabular}[t]{ll}
 \hline
 \textbf{Parameter} & \textbf{Value} \\ [0.5ex] 
 \hline\hline
AFDM frame size, $N$ & 2048 \\ \hline
Carrier frequency, $f_c$ & 79 GHz \\ \hline
Bandwidth, $1/\Delta t$ & 30.72 MHz \\ \hline
Frame duration, $T$ & 66.6 $\mu$s\\ \hline
Maximum range ($\tau_{\max}$) & 98 m (0.65 $\mu$s) \\ \hline
Maximum communication range & 196 m \\ \hline
Maximum speed ($f_{\max}$) & 308 km/h (45 kHz) \\ \hline
$k_f$ & 4\\ \hline
Modulation scheme for communication & QPSK\\ \hline
Number of trials & 100000 \\ \hline 
\end{tabular}
\label{tab:setup}

\end{table}%
Root mean square error (RMSE) computed over several trials is used as the performance metric to evaluate the accuracy of range and velocity estimation.
In each trial, target delays $\tau_i$ and Doppler shifts $f_i$ are generated using a uniform distribution over $\left[0,\tau_{\max}\right]$ and $\left[-f_{\max},f_{\max}\right]$, respectively. Gains $h_i$ are generated via a standard complex Gaussian distribution. The signal-to-noise ratio (SNR) of target $i$ is $\mathrm{SNR}_i \triangleq \frac{\mathbb{E}\{|h_i|^2\}}{N_0}$.
For AFDM, this estimation is done by solving \eqref{eq:max_ML_approx} using the refined-grid search method of Section \ref{sec:estimation}. A multi-pilot version of the same algorithm is adopted for OCDM. As for OTFS, the refined-grid search algorithm detailed in \cite{gaudio2020effectiveness} is employed. All of these algorithms have the same complexity order.

\begin{figure}
    \centering
    \includegraphics[width = .47\textwidth]{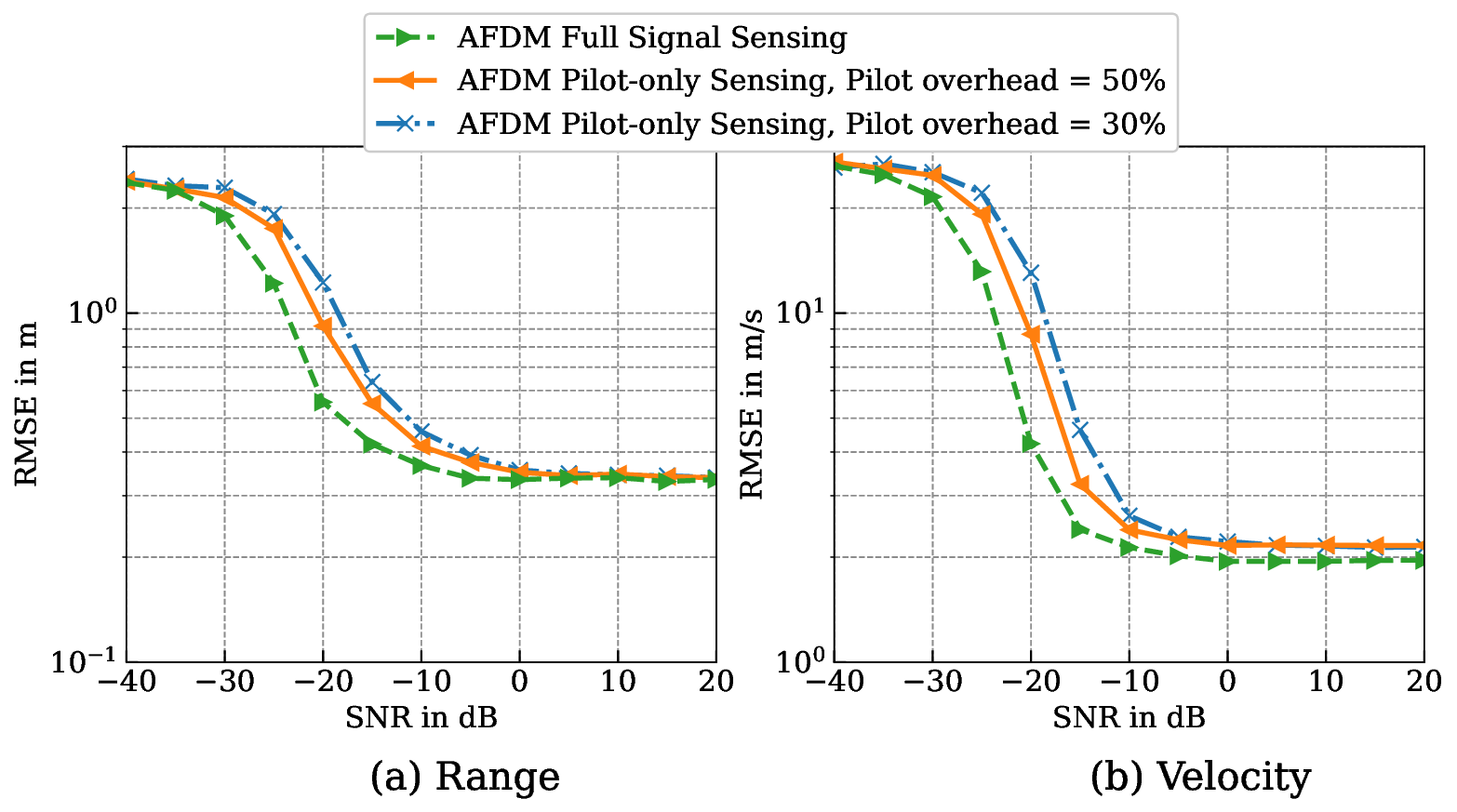}
    \caption{AFDM sensing performance as function of sensing resources}
    \label{fig:RMSE_pilot_total}
    \vspace{-3mm}
\end{figure}
\begin{figure}
    \centering
    \includegraphics[width = .47\textwidth]{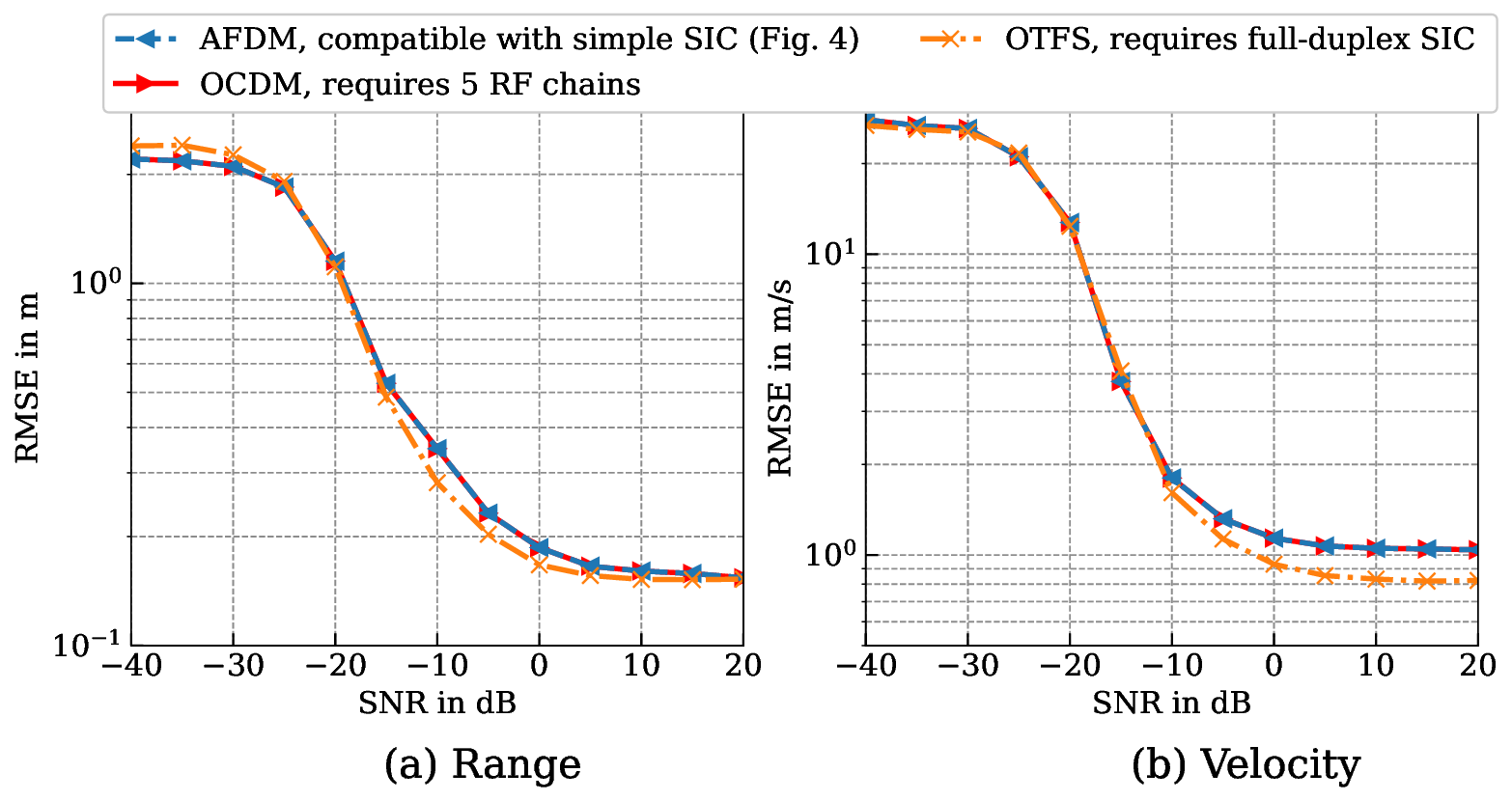}
    \caption{Sensing performance of AFDM, OCDM, and OTFS}
    \label{fig:RMSE_AFDM_OTFS}
\end{figure}
\begin{figure}
    \centering
    \includegraphics[width = .37\textwidth]{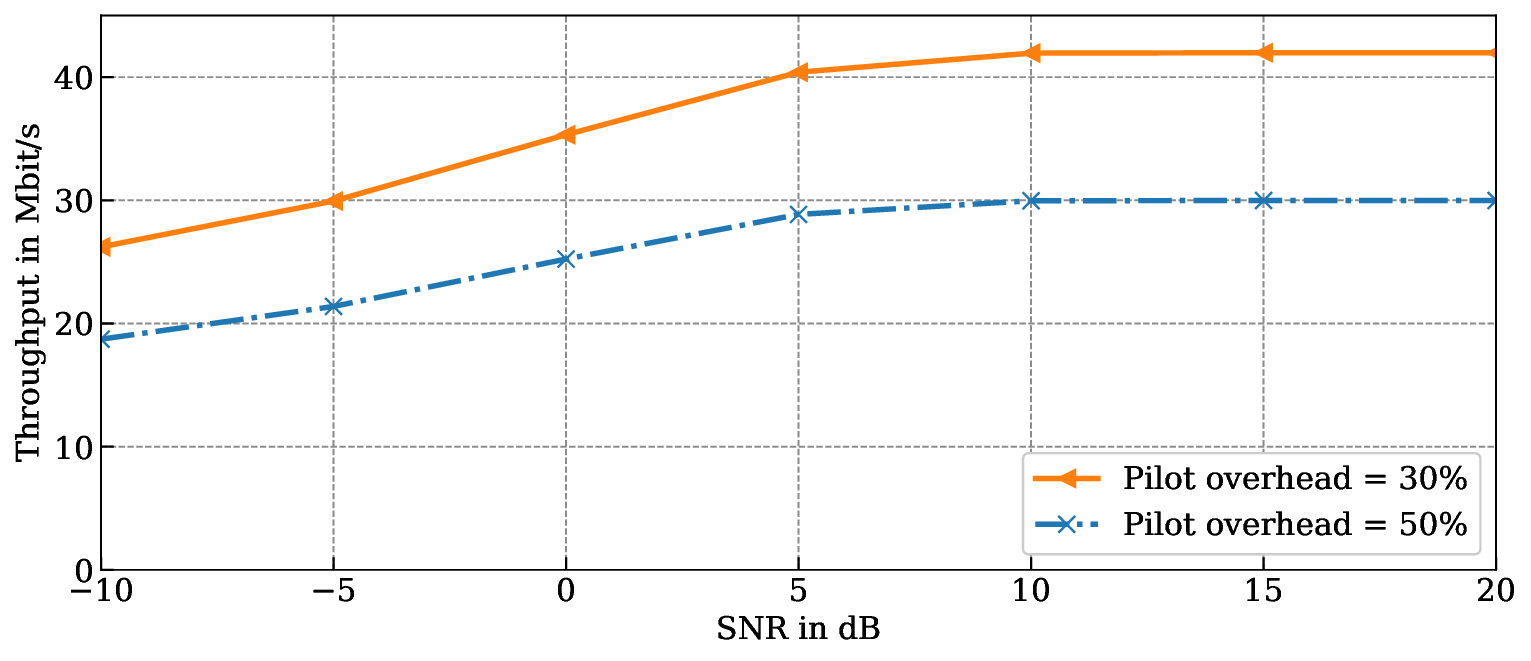}
    \caption{ Throughput of the communications link of Fig. \ref{fig:ISAC_diagram} }
    \label{fig:throughput}
\end{figure}

 In Fig. \ref{fig:RMSE_pilot_total}, the RMSE performance of range and velocity estimation using AFDM with one DAFT domain pilot symbol surrounded with a DAFT domain guard interval is given for two values of the total pilot-guard overhead, namely 30\% and 50\%, and compared to the RMSE performance when estimation is based on the whole AFDM frame. We observe that having a higher pilot guard overhead dedicated to sensing improves RMSE performance (thanks to the boosting of the pilot symbol transmit power a larger guard overhead enables), but has almost no effect on its saturation level, which can be thought of as the effective resolution of range and velocity estimation. This result is in accordance with the fact that one DAFT domain pilot gives the full representation of doubly selective channels.  Another effect of a larger pilot overhead is a reduced communication rate (see Fig. \ref{fig:throughput}) due to the sensing-communication trade-off inherent in pilot-based sensing.
 In Fig. \ref{fig:RMSE_AFDM_OTFS}, we see that OTFS, OCDM, and AFDM have comparable performance in terms of range and velocity estimation RMSE. However, achieving this performance with OTFS and OCDM requires the use of complex SIC, while when using AFDM only a simple receiver architecture is needed.
\section{Conclusion}
A single-antenna AFDM-based ISAC scheme has been proposed in this letter. Our results demonstrate that sensing can be done with only one DAFT domain pilot symbol yet yielding almost the same resolution performance as using the whole frame. Moreover, AFDM chirp nature provides a unique feature that allows for simple SIC when sensing with one pilot symbol, avoiding the need for expensive full-duplex techniques and hardware. 
\bibliographystyle{IEEEtran}
\bibliography{IEEEabrv,Citations}
\end{document}